# Improving generalization of machine learning-identified biomarkers with causal modeling: an investigation into immune receptor diagnostics

Milena Pavlović[1,2,3,*], Ghadi S. Al Hajj[1], Chakravarthi Kanduri[1,3], Johan Pensar[4], Mollie Wood[5,6], Ludvig M. Sollid[2,7], Victor Greiff[7], Geir K. Sandve[1,2,3,*]

[1] Centre for Bioinformatics, Department of Informatics, University of Oslo, Norway
[2] K.G. Jebsen Centre for Coeliac Disease Research, Institute of Clinical Medicine, University of Oslo, Norway
[3] UiO:RealArt Convergence Environment, University of Oslo, Norway
[4] Department of Mathematics, University of Oslo, Norway
[5] Department of Pharmacy, University of Oslo, Norway
[6] Department of Epidemiology, Gillings School of Global Public Health, University of North Carolina at Chapel Hill, USA
[7] Department of Immunology, University of Oslo and Oslo University Hospital, Norway

* correspondence: geirksa@uio.no, milenpa@uio.no

**Abstract**

Machine learning is increasingly used to discover diagnostic and prognostic biomarkers from high-dimensional molecular data. However, a variety of factors related to experimental design may affect the ability to learn generalizable and clinically applicable diagnostics. Here, we argue that a causal perspective improves the identification of these challenges and formalizes their relation to the robustness and generalization of machine learning-based diagnostics. To make for a concrete discussion, we focus on a specific, recently established high-dimensional biomarker – adaptive immune receptor repertoires (AIRRs). Through simulations, we illustrate how major biological and experimental factors of the AIRR domain may influence the learned biomarkers. In conclusion, we argue that causal modeling improves machine learning-based biomarker robustness by identifying stable relations between variables and by guiding the adjustment of the relations and variables that vary between populations.

# Introduction

High-throughput sequencing technologies now allow for the examination of a variety of patient characteristics, such as genetic variation[1], DNA methylation[2], gene expression[3], gut microbiota[4], and adaptive immune receptor repertoires (AIRRs)[5,6]. Proof-of-concept studies showed that such molecular and biological markers (biomarkers), defined as objective indications of the medical state that can be accurately and reproducibly measured[7], hold great promise for disease diagnostics, especially in combination with machine learning (ML)[2,3,5,8]. However, there exist several challenges to using ML for diagnostics. First, the data used in diagnostic studies may be selected based on availability, e.g., collected from patients visiting the clinic or having a similar genetic background (sometimes referred to as "convenience sampling"). Furthermore, rather than originating from a single source, the data might instead be collected at multiple locations or at distinct time points. These factors may introduce systematic differences between datasets, such as measurement errors and batch effects, which need to be considered when designing a new study, or adjusted for when the data are already collected. A failure to do so can introduce selection and confounding biases that lead to models failing in real-world applications despite showing promising performance during diagnostic development[9–12]. Finally, molecular biomarker data are typically high-dimensional, which makes it more challenging to disentangle noise and biases from the true markers associated with the disease[13].

In ML-based diagnostics, these challenges are examined from two perspectives. One perspective is purely statistical[14,15]: it attempts to solve the challenges by anticipating how the distributions of features or labels will change (a phenomenon called dataset shift) but does not consider causal relations between them, as discussed by Whalen and colleagues in a genomics setting[11]. An alternative perspective investigates these challenges using the causal inference framework[16–18], describing dataset shift using formal definitions with respect to a proposed causal model of the underlying process[19].

The causal inference framework described by Pearl, known as do-calculus[16], can be used to estimate causal effects from non-experimental data whenever the effect is identifiable under a given causal structure. The causal structure can be encoded using a (directed acyclic) causal graph. The nodes in the graph represent the variables in the considered system and the directed edges between the nodes represent direct causal relationships. The do-calculus framework is fully non-parametric, relying only on the causal assumptions implied by the graph, which is typically based on domain knowledge (or more recently, through causal discovery – learning the causal graph from structured data, and causal representation learning – learning the causal variables from the data[19]). Of particular importance for this work, the graph-based approach to causal analysis has also been shown to be a highly useful tool for the task of data fusion where multiple datasets, sampled under heterogeneous conditions, are combined to answer a probabilistic (and possibly causal) query of interest, thus dealing with biases that emerge due to environment change, confounding, and sample selection. For an introduction to this field, see the work by Bareinboim and Pearl[20], and for a brief synopsis of concepts, see Focus Box 1.

Causal inference is increasingly seen as a key component in diagnostics[21], medical image analysis[10], decision-making in healthcare[22], clinical risk prediction[23], and the clinic in general[24], to replace purely associative machine learning models with models that explicitly consider the data generating process. When applying ML in populations where disease prevalence changes, or where populations differ in the distribution of certain features like age, the model performance in the new setting may be substantially different. The causal model of the biological process may be used to determine which mechanisms (conditional distributions of variables of interest given their direct causes) will remain stable across populations. The ML model may then be based on these invariant (stable) mechanisms[19] ensuring the robustness of the ML model in the presence of different dataset shifts[9,15,25,26]. In addition to potentially improving the performance of learned models, causal considerations can also help to formally or intuitively reason about how diagnostic accuracy may be affected by a variety of differences between application contexts.

a  Overview of an AIRR diagnostic development pipeline

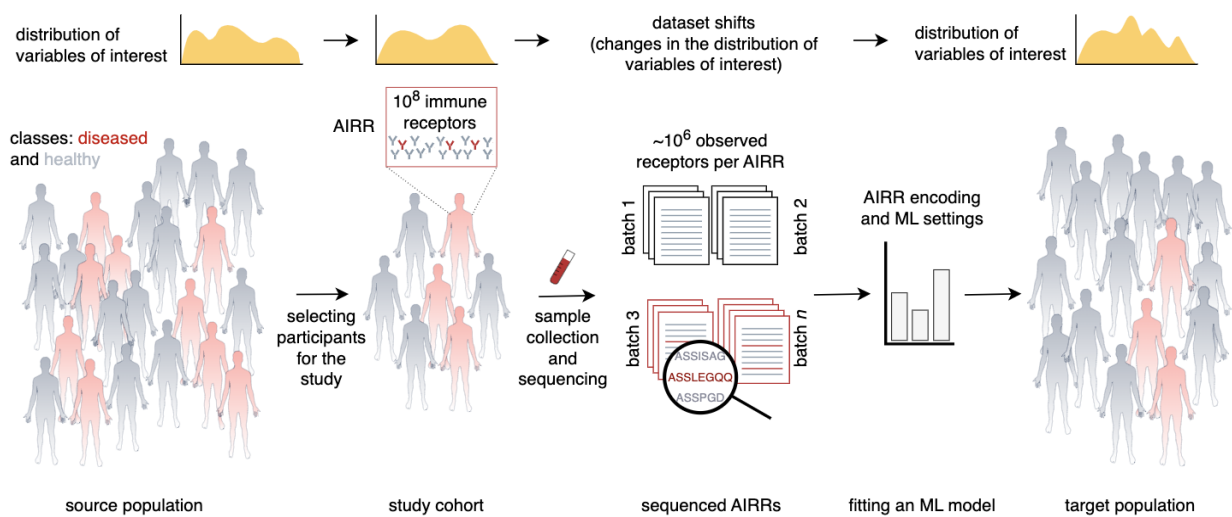

b  An example of a causal graph in the AIRR domain

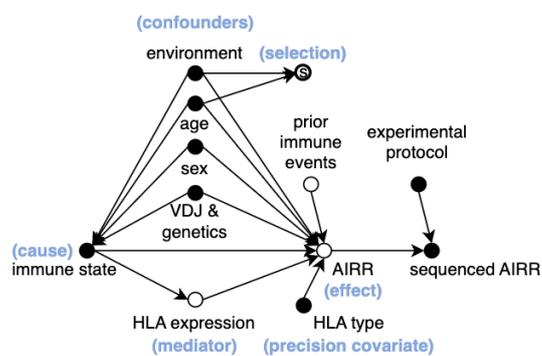

c  Causal and anticausal prediction

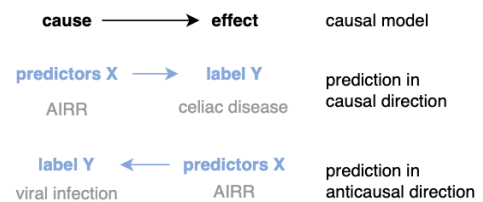

**Figure 1**: **Developing an AIRR-based diagnostic. a.** Overview of the diagnostic pipeline based on adaptive immune receptor repertoires (AIRRs), including patient collection from the source population, sampling, sequencing with batch effects, ML method development, and application in the target population. **b.** An example of a causal structure of AIRRs and the immune state where the nodes represent different variables involved (repertoire, HLA type, age, immune state) and the arrows represent causal relationships between variables. Solid nodes in the graph (immune state, HLA type, sequenced AIRR) denote observed variables, while empty nodes are not observed (HLA expression, prior immune state). The node with S inside is the selection node with edges showing what variables influenced the selection of participants for the study. **c.** For AIRR-based diagnostics, predictions may be either made in the causal direction (predicting the effect from the cause, e.g., in autoimmunity), or in the anticausal direction (predicting the cause from the effect, e.g., in infections) making the models predicting in the anticausal direction potentially less stable since they are not modeling the biological mechanism.

> **Focus Box 1.** A brief introduction to causal inference
>
> Causal inference research aims to estimate causal effects between variables of interest, typically by introducing certain assumptions regarding how the variables relate to each other in a causal way. We say that variable C has a causal effect on variable E if intervening to change variable C would change (the distribution of) variable E[16].
>
> Here we briefly define basic concepts and different types of variables important in the causal inference field. For a more detailed introduction to the field, see the review by Bareinboim and Pearl[20], and for connections to ML, see the works by Schölkopf and colleagues[19,27].
>
> **A structural causal model** consists of a set of variables of interest and a set of functions that describe how the values of the variables are assigned and describe their dependencies on other variables. Such models describe (often partially) the data-generating process. The causal structure of the model can be represented by the **causal graph** (Figure 1b), where the nodes represent variables, and each edge defines the influence of one variable on another. The absence of an edge between two nodes implies that there is no direct causal relation between them.
>
> Assuming a causal graph structure, typically in the form of a directed acyclic graph, the do-calculus framework[16] can be used to identify if it is possible to estimate the causal effect between two variables C (cause) and E (effect) from the available data under the given assumptions (Figure 1b), and it specifies how to estimate it when possible. Other variables might influence the effect estimation[20]: **confounders** are variables that causally affect both C and E (C ← confounder → E), **colliders** are influenced by both C and E (C → collider ← E), **mediators**[28] are intermediary variables between C and E (C → mediator → E), that describe the mechanisms of how C influences E indirectly. **Moderators** (effect modifiers[17]) change the relation between C and E depending on the moderators' values[29]. **Precision covariates**[30] are variables that influence only C or only E and may be used to improve the precision of the estimators in some cases.
>
> Causal graphs can be arbitrarily complex, and variables may not have as clear roles as described above. To enable consistent estimation of the causal effect it is necessary to analyze the paths in the graph to prevent bias. A particularly useful technique for preventing bias is known as the **backdoor criterion**, where the idea is to close all non-causal paths with incoming arrows into both C and E (backdoor paths) while simultaneously keeping all directed (causal) paths from C to E open. Briefly, a path is defined as open if every collider on the path (or a descendant of the collider) is controlled for and any other variable on the path is not controlled for. Controlling for a confounder (e.g., age, Figure 1b) is a simple example of closing a backdoor path. Cinelli and colleagues provide an overview of good and bad controls[31] in the context of estimating causal effects.

Dataset shifts are especially important to handle properly when using ML in healthcare to avoid incorrect diagnosis or a suboptimal course of treatment[32], and causal inference can help with that. Specifically, when analyzing sequencing data to discover molecular biomarkers of disease, there are two possible underlying causal structures assuming there is a causal relationship between the biomarkers and the disease. The biomarkers represented by the sequencing data may be causing the disease or may arise as an effect of the disease. This means that the diagnostic prediction may be either in a causal direction (predicting the effect from causes, e.g., finding changes in the sequencing data that have played a role in causing a disease), or in an anticausal direction[33] (predicting causes from effects, e.g., finding changes in the sequencing data that occurred as a consequence of the disease, Figure 1c). Depending on this direction, dataset shifts will manifest in different ways. For example, when predicting in the causal direction ($X \rightarrow Y$), we might expect the performance to be more stable under changes to $P(X)$ since our target, $P(Y|X)$, is a component in the causal factorization $P(X, Y)=P(X)P(Y|X)$ and thus independent of $P(X)$ due to the principle of independent causal mechanisms[19]. On the other, we would expect no such robustness under the anticausal direction ($X \leftarrow Y$) since $P(Y|X)$ is now a component in an entangled factorization that does not follow the causal structure.

Due to the opportunities for ML in this relatively new field, and in the interest of a concrete discussion, we here focus on adaptive immune receptor repertoires (AIRRs), which are increasingly used for diagnostic purposes[5,6]. AIRRs are high-dimensional molecular markers reflecting past and present immune responses of a patient and can be efficiently assayed based on targeted high-throughput sequencing from a standard blood sample (Figure 1a, Focus Box 2). AIRR-based approaches may enable earlier diagnosis and prognosis, complement existing diagnostic tests, and have in principle the capacity to diagnose a broad range of diseases by a single test[5]. For this reason, AIRRs are highly promising as biomarkers of immune-mediated diseases[5,6]. Proof-of-concept studies show that statistical and ML approaches on AIRRs could be used to diagnose or predict the outcome of for example different types of cancer[34,35], celiac disease[36,37], multiple sclerosis[38], rheumatoid arthritis[39], systemic lupus erythematosus[39], cytomegalovirus[40] and hepatitis C virus[41]. However, further validation studies on external cohorts are needed to establish the robustness of trained models to batch effects, preferential participant selection, or confounders like age and sex. Here, we discuss why and how accounting for the biological and experimental aspects of the underlying data-generating processes is crucial for clinically relevant diagnostic development. Our proposed approach is based on making modeling assumptions explicit and relying on best practices in machine learning and causal inference to improve the generalizability of results.

## Challenges in AIRR diagnostics study design

The main challenge of ML in diagnostic settings is whether the probability distributions learned from the training data (e.g., AIRRs) will generalize to new application settings (e.g., clinical settings). We discuss this challenge in more detail in this section.

To define the problem, the set of examples (here study participants) available at training time will be called the *study sample* (or *study cohort*), sampled from the underlying *source population* (Figure 1a). The population on which the classifier will be applied is called the *target population*. In addition to *population*, terms often appearing in this context are *environment* and *domain*. Source and target populations are also referred to as *development* and *deployment populations* or *domains*. If the source and target populations have the same joint probability distribution (disease prevalence and feature distribution both stay the same, as well as the relations between them), the estimated ML model can be readily applied, provided that it is internally valid (Focus Box 3). This means that when the probability distribution is stable, a classifier trained on the source population can be directly applied to the target population, with the expected performance being the same as for the source population. This performance is estimated on the test dataset – the portion of the original study cohort that was used neither for training nor for selecting between different ML approaches[42]. This represents the estimated performance when applied to new, independent, and identically distributed (i.i.d.) examples (e.g., AIRRs), i.e., when the source and target population have the same distribution. Even when representative, the training and test sets (e.g., coming from the study cohort) are still finite samples of the underlying population, and the variability inherent in this sampling process leads to uncertainty in the estimated performance. In the i.i.d. setting, this uncertainty can however be quantified, e.g., through resampling procedures such as bootstrapping.

**Focus Box 2**. Adaptive immune receptor repertoires (AIRRs)

Adaptive immune receptors (AIR) are proteins created by B and T cells that specifically recognize parts of foreign or self-antigens, such as viruses or cancerous cells, and mount an immune response to neutralize them[49,50]. AIRs are highly diverse, with an estimated $10^{15}$ different receptors[51–54], and specific to certain antigens. Their diversity arises from a stochastic process called V(D)J recombination that combines V, (D), and J gene segments with random insertions and deletions to create receptors able to detect a variety of antigens an individual encounters in their lifetime[55–57]. The most important region of an AIR for recognizing an antigen is called complementarity-determining region 3 (CDR3)[58,59]. It is an, on average, 15 amino acid long part of the receptor with the highest variability. When examining antigen recognition and binding, it is most often the only part of the receptor that is used in computational AIRR analyses[49,60,61].

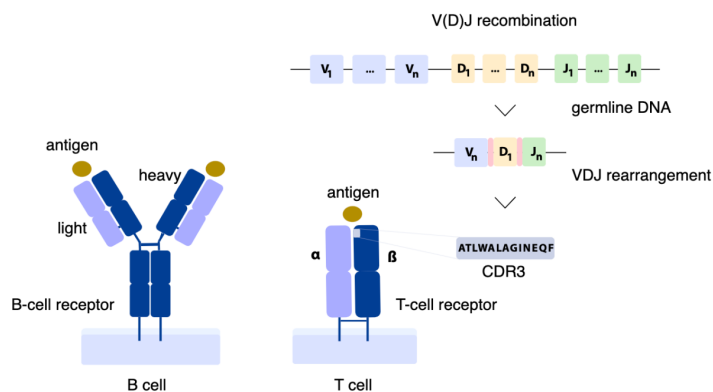
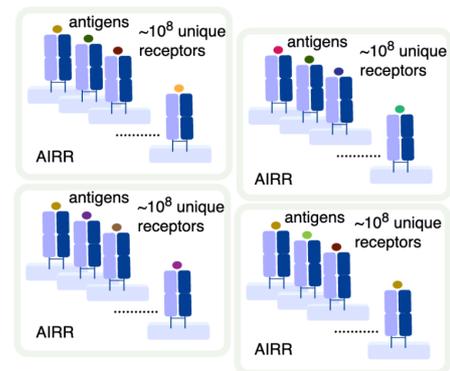

Adaptive immune receptor repertoires (AIRRs) are sets of all AIRs present in an individual. There are an estimated $10^8$ unique receptors with frequency distribution specific to an individual[62–64] at any given time, with very few receptors specific to one antigen. Examining the AIRs in AIRRs provides unique insights into disease, rendering AIRRs a major target of current diagnostic biomarker research[5]. This task is challenging due to the low overlap of receptors between AIRRs of different individuals[65], and the unknown specificity of individual AIRs determined by complex sequence patterns. This inspires the ML applications for AIRR-based diagnostics[6].

While statistically convenient, the i.i.d. assumption rarely holds in the real world[43–45] — the probability distribution might change from source to the target population in the marginal or conditional distribution of variables[25,46]. Marginal distributions may change due to label shift (e.g., change in disease prevalence) or covariate shift (e.g., change in age distribution). The conditional distribution of variables may change if it describes an anticausal relation (when predicting the cause from the effect, e.g., immune state from AIRR, Figure 1c) or due to the occurrence of unstable mechanisms[9] (e.g., changing the time of sequencing in the course of the disease might result in estimates that only hold for the study cohort). Importantly, these shifts reflect systematic biases that would hold up even if a study cohort was infinitely large, and their extent cannot be quantified based only on information from the source population. The biases may arise from different aspects of the data-generating process and are formally defined with respect to their source (leading to the identification problem as defined in the causal inference)[17]. This is discussed in more detail in the next sections.

> **Focus Box 3**. Internal and external validity for ML classification
> 
> **Internal validity.** In the field of causality, a conclusion of a scientific study is said to be internally valid if it is true of the population on which the study was conducted[73]. In the ML sense, we define internal (in-distribution) validity to be based on learning the source population distribution instead of noise and to be assessed by procedures such as cross-validation, leave-out test dataset, or bootstrapping[42]. Failure to comply with this requirement leads to a model that is overfitted to the data and is thus overly optimistic. The correct procedure ensures that the obtained performance estimate reflects how the model is expected to behave when applied to the new data coming from the same distribution. In the ML literature, the performance on new data independently sampled from the same distribution is referred to as generalization[74]. Varoquaux and Cheplygina[75] and Walsh, Fishman, and colleagues[42] provide recommendations for best practices in ML for biology and medicine.
>
> **External validity.** External validity (also called transportability) is in the field of causality defined as the ability to generalize results to new environments or populations[76]. In the ML field, external validity is typically examined in the context of domain adaptation[15], domain generalization[77], and out-of-distribution generalization subject to certain constraints (e.g., as described by Nagarajan and colleagues[78]). External validity as defined in the field of causality is indeed often the main aim of scientific analyses, in that external validity is typically achieved through the discovery of invariant mechanisms across environments.

To build an AIRR-based diagnostic, the typical workflow is as follows: first, the study cohort is selected from the underlying source population. The blood or tissue samples are then obtained from disease-affected (cases) and healthy individuals (controls) and the targeted cell population (e.g., T cells) are sequenced via bulk or single-cell sequencing, obtaining a set of approximately $10^5$–$10^6$ immune receptors per individual that can be used to learn the patterns indicative of disease (Figure 1a). Such an approach needs to take the following challenges into account: (i) the individuals included in the study represent only a small part of the underlying population (leading to sampling variability, as defined previously, and discussed by Bonaguro and colleagues[47] and Raybould and colleagues[48]). (ii) The individuals included in the study do not necessarily have the same distribution of variables of interest (e.g., age) as the population where the classifier will be applied due to preferential selection of participants (selection bias), (iii) some of those differing variables might be connected to both AIRR and the immune state, introducing spurious correlations. Finally, (iv) there might exist measurement errors in the AIRR data due to, for example, the PCR-based library preparation or sequencing process, which in some cases may be dependent on other variables and potentially introduce spurious correlations. Experimental settings influencing the observed properties of AIRR data on the example of B-cells are discussed in more detail by Raybould and colleagues[48].

To illustrate these concerns, we introduce an example of building an AIRR-based diagnostic for a viral infection (Figure 1b). In this example, the immune state is defined as the presence of the pathogen of interest in an individual in a way that gives rise to changes in AIRR. In addition to the immune state, AIRR is also influenced by prior immune events (e.g., prior infections or vaccinations), age[66], sex[67], genetics (including the V(D)J recombination model[57]), and the environment. The environment here refers to the geographical location and may be a proxy for socioeconomic conditions. The immune state may also influence the AIRR through (typically unmeasured) human leukocyte antigen (HLA) expression, such as in EBV[68]. The HLA type influences the AIRR as well[69,70]. Finally, the observed sequencing data is not a perfect reflection of an individual's AIRR. Observed AIRR data reflects only a limited proportion of a patient's full AIRR, introducing additional sampling variability. Also, the experimental protocol may introduce systematic biases in terms of which receptors are captured[71,72], which is especially problematic if the experimental protocol varies in a way that correlates with other patient variables of interest (covered in more detail in the later sections). For other types of diseases, such as autoimmunity or cancer, the causal graphs might differ.

## Confounding: how age, sex, and environment affect the analysis

Age, sex, and genetic background have been demonstrated to influence the immune repertoire. The repertoire diversity decreases with age[66,79], and sex influences the usage of V genes in T-cell receptor repertoires[67]. Sex and age also influence the immune state for example through their influence on innate immunity[80,81], and therefore represent potential confounders in disease diagnostics (Focus Box 1). Environment, here broadly defined as a proxy for socioeconomic background and geographic location, may also act as a confounder, but further studies are needed to explain how exactly it influences the immune state and AIRR.

For predictive purposes, confounding is in general not problematic, unlike selection and measurement biases[82]. However, if the source and target populations differ, confounder distributions (or even functional relations due to e.g., differences in measurement policies) may change, potentially changing the perceived (non-causal) relationship between the immune state and AIRRs. Correa and Bareinboim formally describe when and how the classifier may be transferred to the new, target population, referring to this as statistical transportability[83]. Magliacane[84], Subbaswamy[85], and Rojas-Carulla and colleagues[86], on the other hand, propose different methods to select a subset of features so that relations remain invariant across the source and target population, even in cases of partially known causal models.

For purely diagnostic purposes, stable confounders are not problematic, and might even improve the predictive performance of a model. However, one should note that the recovered biomarkers could be reflecting the confounders just as much as the immune state[11]. If the aim is to obtain biological insight (or discover causal effects), confounding should thus be controlled for.

Additionally, it is important to ensure the study participants are representative of the population in terms of their age[66,81] and sex[67,80]. For example, a diagnostic may be developed using a predominantly male study cohort but applied to the population where sex distribution is different, changing the performance of the learned model. For more details on selecting a study cohort, see the section on *Selection bias: choosing participants for the study*.

In Experiment 1, we illustrate the effect of a confounder on predictive performance and discovered markers, and we show that the performance may depend on the stability of the confounder distribution and the strength of the confounder's influence on AIRRs.

## Timing of measurements, measurement bias, and batch effects

Batch effects are systematic biases in the statistical sense, connected to experimental protocols exhibiting different behavior across conditions[87]. A certain level of bias is always present in sequencing (molecular biomarker) data[87]. If the batch effects are independent of the labels to be predicted, this will not introduce any bias in the learned ML models, leading only to increased variability of the biomarker, with reduced data efficiency as consequence. In this case, the sequencing protocol behaves as a precision covariate in the causal model: it does not bias the resulting predictor but can influence the precision of the predictor. Batch effects are more problematic when correlated with the label (e.g., immune state), in which case a predictive model may achieve good performance in the study cohort by learning batch effect associations. Such a model would fail when applied to the source or target population, where the batch effect would not be associated with the disease or would be associated differently. This would be the case regardless of the direction of the underlying causal structure (biomarkers causing the disease or the disease causing the biomarkers). Batch effects in the immunomics setting are further discussed by Barennes and colleagues[71] and Bonaguro and colleagues[47]. An illustration of these effects is provided in Experiment 2 (Figure 4c, d).

Closely related to batch effects is the measurement bias that occurs when the association between the cause and effect in question (biomarkers and label) is changed by the data measurement process[17]. It can happen through various causal mechanisms and may in the worst case make learning from the data impossible. For example, the AIRR sequencing process is often subject to sequencing errors[71]. If there are a lot of errors or if spurious correlations occur due to the measurement process, the errors cannot be corrected for since the ground truth (e.g., exact receptor sequences) is not known. Additionally, if the study participants are treated differently based on their known immune state, e.g., by being enrolled by different institutions or at different time points, the observed repertoire then might not represent the underlying repertoire sufficiently well (Figure 1b). This is also of interest when multiple datasets (e.g., available through public repositories, such as AIRR Data Commons[88]) need to be combined for a study that might include distinct measurement biases. Therefore, measurement bias poses challenges to both estimating the ML model on the study cohort, as well as applying the model to the target population.

The timing of measurement, i.e., in the AIRR example, when the sequencing is performed in the course of the disease, is also very important for diagnostic development. The performance of the diagnostic depends on how well the model was able to capture the underlying signal and how well the source population represents the target population where the model will be deployed. For example, the positive examples (diseased individuals) in the dataset used for training might be collected retrospectively, after individuals exhibited symptoms of the disease and were diagnosed and possibly already treated by medical professionals. In this case, the collected AIRRs will not be representative of the AIRRs of individuals who would get tested to establish the diagnosis. To mitigate this issue, the study cohort should be representative of the target population in terms of the timing of measurement or at least include individuals sequenced across the disease progression spectrum.

## Selection bias: choosing participants for a study

To develop a diagnostic of a viral infection, information must be obtained from individuals who were tested for the virus with both positive and negative results. If only individuals with symptoms were tested, the developed diagnostic would be biased. This bias would occur because the study cohort was not representative of the source population with respect to variables of interest (immune state, AIRR), being based only on the individuals exhibiting the symptoms and ignoring potentially asymptomatic individuals. For more details, see the work by Griffith and colleagues discussing a similar example for COVID-19[89].

The preferential selection of participants for the study gives rise to selection bias. In the causal inference literature, selection bias is broadly defined as any statistical association that is a consequence of selective inclusion into the study cohort (e.g., participants are recruited based on some of the variables of interest in the analysis)[90,91]. It does not depend on the cohort size[92] and can introduce, increase, decrease, or even reverse the sign of the existing associations[93].

Given a directed acyclic graph where the nodes are variables of interest and directed edges show which variable causes other(s) (Figure 1b, Focus Box 1), preferential selection might arise when selecting based on: (i) the common effect (collider) of the cause variable and the effect variable (Figure 2a), or (ii) when selecting based on the value of a confounder (Figure 2b) or (iii) some other non-collider variable[91] (Figure 2c). For the example in Figure 1b, selection bias can be introduced if the inclusion of participants was influenced by the confounder, such as age. This means that selection bias may be present whenever the data collection process depends on the cause and effect or their parents in the causal graph[17] (e.g., Figure 1b). Controlling for a confounder resolves the selection bias in this case. We illustrate the influence of selection bias on the predictive performance of ML models in Experiment 2.

In ML, selection bias has a much broader and less structural definition: selection bias occurs when there is a difference in the marginal distribution of any variable used for prediction (covariate shift)[46] between the study cohort and the source population or a difference in the marginal distribution of the label (such as the disease status, resulting in label shift[94]). These definitions do not always rely on causal diagrams, but we argue that causal diagrams and the underlying causal models can improve ML analysis by providing guidance on how to recover from biases under a set of assumptions.

Closely related to the selection bias issue is the concept of transportability[20,76] (also called external validity, Focus Box 3). Selection bias with respect to the causal model exists if the participants in the study differ from the source population in a way that is influenced by the cause, effect, or their parents in the causal model. Transportability bias, on the other hand, occurs when we move from a well-defined source population to a distinct target population, where the target and source populations are at least partially non-overlapping[95]. An example might be an AIRR-based diagnostic built in Norway (source population) that needs to be applied in Serbia, where Serbia could represent a target population that might arbitrarily differ in the marginal distribution of variables, such as HLA or age, but for instance, the causal mechanisms of the disease (the influence of the pathogen on AIRRs given HLA and age) remain stable.

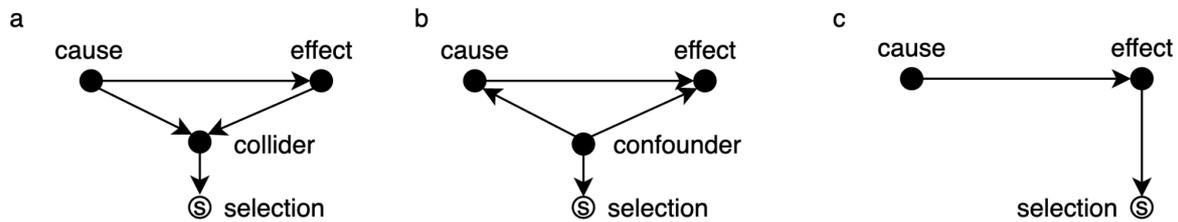

**Figure 2**: **Examples of selection bias in causal models.** Selection bias may occur **a**. by selecting based on a collider, **b**. because a confounder influences selection, or **c**. by selecting based on the effect variable.

# The same biological variable can take on alternative causal roles: HLA and antigen presentation

Some biological variables can have different roles in the causal graph depending on the disease or immune event, with important implications for the analysis. One such example is HLA. For T cells to recognize pathogens and mount an immune response, peptide fragments derived from pathogen proteins have to be presented on the cell surface bound to HLA molecules. The complex of peptide and HLA is recognized by the T-cell receptor, and upon recognition, the T cell will be activated to exert effector functions and undergo T-cell clonal amplification. While having a critical role in presenting pathogen-derived peptides to T cells, the HLA molecules also shape the T-cell receptor repertoire during positive and negative selection during T-cell maturation in the thymus[96]. HLA can thus affect the T-cell receptor repertoire composition of both naive and antigen-experienced T cells. This makes HLA an important variable for the development of diagnostic tests.

To discuss the role of HLA in disease, we focus on the HLA allotype, defined as the presence of specific genes and alleles. Depending on the assumptions of the causal model describing the disease (Focus Box 1), the role of HLA in the analysis will differ. For a viral infection, such as discussed by Emerson and colleagues[40], HLA type is what is sometimes referred to as a precision covariate (Figure 3a): it will influence the AIRR, but not the immune state. Theoretically, adjusting for it will not resolve any bias, but might improve the precision of the diagnostic. In practice, this might depend on how much data are available for different HLA allotypes.

In AIRR-mediated autoimmunity, however, where AIRR causes the immune state, HLA might influence both the immune state and AIRR and have a role of a confounder (Figure 3b). Additionally, HLA can be a moderator (Focus Box 1), for example in celiac disease. Celiac disease is an autoimmune condition that can occur because of gluten consumption in subjects who carry HLA-DQ2 or HLA-DQ8. The presence of HLA-DQ2 or HLA-DQ8 is then a necessary condition for celiac disease. Other HLA allotypes, even if the person in their AIRR represents T-cell receptors having hallmarks of gluten-specific T-cell receptors, will not result in disease as these HLA molecules are unfit to present the culprit gluten epitopes[97]. The presence of these allotypes is, however, not a sufficient condition for the disease. Therefore, exploiting HLA information in the analysis might then be beneficial and even necessary to distinguish between AIRRs of healthy individuals with or without an HLA allotype necessary for the disease versus AIRRs of individuals who have developed the disease.

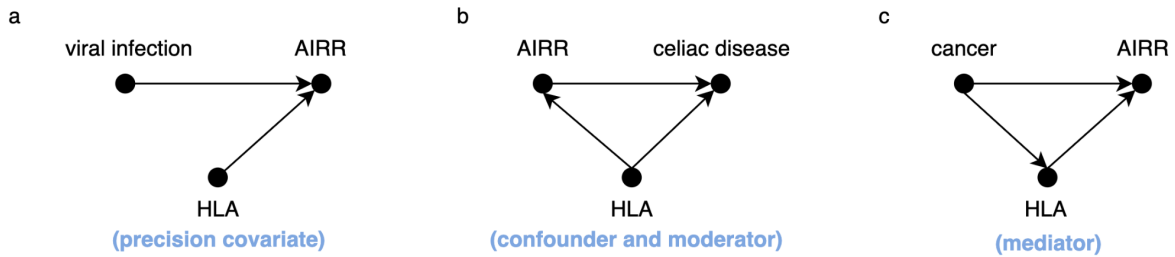

Figure 3: **Different causal roles HLA can take for different types of immune-related diseases. a**. In a viral infection[40], HLA is a precision covariate. **b**. In celiac disease[97], HLA is both a confounder and a moderator. **c**. In cancer[98], HLA can be a mediator.

In some cancers, tumor cells have somatic mutations that affect peptides binding to HLA and help tumor cells evade immune recognition[98]. In this case, HLA acts as a mediator between disease and the AIRR (Figure 3c), and as such does not need to be adjusted for in the analysis when developing a diagnostic.

## High dimensionality of AIRR data

Building diagnostics based on AIRRs (and molecular data in general) is made more challenging by the high dimensionality of the data. We have in this paper so far represented an AIRR by a single node (variable) in the causal graph, which then represents millions of individual AIRs.

The first question in this setting is how the AIRRs should be represented. A complete AIRR representation may be a set of all possible immune receptors annotated with their specificities and information if they are present and at what frequency in an AIRR of an individual. This representation is not computationally feasible, as there might be more than $10^{15}$ such receptors[51–54], and receptor specificities are largely unknown. Alternative ways to represent an AIRR might be by (i) a set of immune receptors as provided by the sequencing procedures, (ii) a set of immune receptors relevant for the immune state provided that it is possible to determine the relevant ones, (iii) a summary statistic of receptor characteristics, such as parameters of the distribution describing the number of immune state-specific receptors with respect to the total number of receptors in an AIRR, or k-mer frequencies of AIRRs, (iv) through physicochemical characteristics of the receptors. (v) The AIRR variable in the causal model (Figure 1b) could also be expanded to include the details of VDJ recombination describing the data generation process. (vi) Finally, the latent AIRR representation could be learned from the data, e.g., via autoencoders or similar methods coupled with domain-specific restrictions, although interpreting the latent variables and mapping them back to the causal model may be a challenging task.

Provided that a useful representation is chosen, a causal model with many causes or many effects (depending on whether AIRRs cause the immune state for the specific disease or the opposite) is still an active area of research. Some work focuses on the setting where causes are high-dimensional, for example when estimating the effect of gene knockdowns in yeast[99]. Other focus on high-dimensional confounders[100,101]. Castro and colleagues discuss the causality for machine learning in the high-dimensional setting of medical imaging[10]. However, while there are parallels to imaging, the imaging causal models differ significantly from AIRRs and molecular biomarkers more broadly, posing the question of exactly how the different biases discussed above manifest in high-dimensional sequencing data. Their influences are different: diversity changes, skewed gene usage, batch effects, and varying generation probabilities of individual receptors. Handling these biases in high-dimensional settings of AIRRs remains an open problem.

So far, we used AIRRs to refer to both single and paired-chain T-cell and B-cell receptor repertoires, although often only single-chain T- or B-cell receptors are sequenced and partially observed. The causal model could be extended to include paired chain T-cell and B-cell receptor repertoire data as separate variables depending on the research question. Each of these adaptive immune receptor populations might be further split for causal modeling to allow for complex interactions between different cell subtypes[102] or localizations[103].

# Exemplification of how the underlying causality may influence the performance of predictive models: a simulation study for the AIRR diagnostics setting

To illustrate the influence of different variables in the causal model on the performance of ML algorithms for diagnostics, we performed three simulation experiments where we systematically varied causal parameters. In the first two experiments, we trained a model to predict the immune state based on AIRR data without taking potential confounders (e.g., age or sex) into account (Figure 4). In the third experiment, we contrasted the handling of batch effects for the AIR setting versus a different molecular biomarker where batch effect handling is more established. The code for the experiments is available on GitHub (https://github.com/uio-bmi/CausalAIRR) and the results are deposited to Zenodo[104–107].

## Effects of confounding and study selection bias in AIRR settings

The first experiment illustrates how the presence of confounding influences the prediction of the immune state for different strengths of confounding (Figure 4a, b). In this experiment, the confounder and immune state are binary variables: the confounder has values *C1* or *C2* and the immune state can be *diseased* (positive) or *healthy* (negative). The parameters of the probability distribution of the immune state depend on the value of the confounder, making examples with confounder value *C1* much more likely to be *diseased*. We constrain the resulting distribution of the immune state variable to have balanced classes (approximately the same number of *diseased* and *healthy* examples), following the typical setting of balanced sampling in ML, and then vary the confounder distribution while keeping the classes balanced to examine the effect of the confounder on prediction performance. As discussed in the section *Confounding: how age, sex, and environment affect the analysis*, the presence of a confounder is not always an issue for the prediction task: if the confounder distribution does not change from source to target population (from training (validation) to testing), similar performance can be expected on both datasets (Figure 4b). However, if the confounder distribution changes (the parameter of the binomial distribution), the performance may drop substantially (Figure 4b).

In the second experiment (Figure 4c), we examine how the presence of selection bias and batch effects influences immune state prediction. The causal graph includes (i) the immune state which causes changes in AIRRs, (ii) the hospital the patients came from, (iii) the experimental protocol for sequencing AIRRs determined by the hospital, and (iv) the sequenced AIRRs themselves. We simulate the training dataset in the presence of selection bias, which introduces a correlation between the hospital and the immune state, even though the hospital does not have any influence on the immune state. When an ML approach is trained on data that are biased in this way, it also learns the signal of the sequencing protocol since it is predictive of the immune state through the spurious correlation between the immune state and the hospital. When such a model is then applied to new data that are not biased, its performance decreases (Figure 4d).

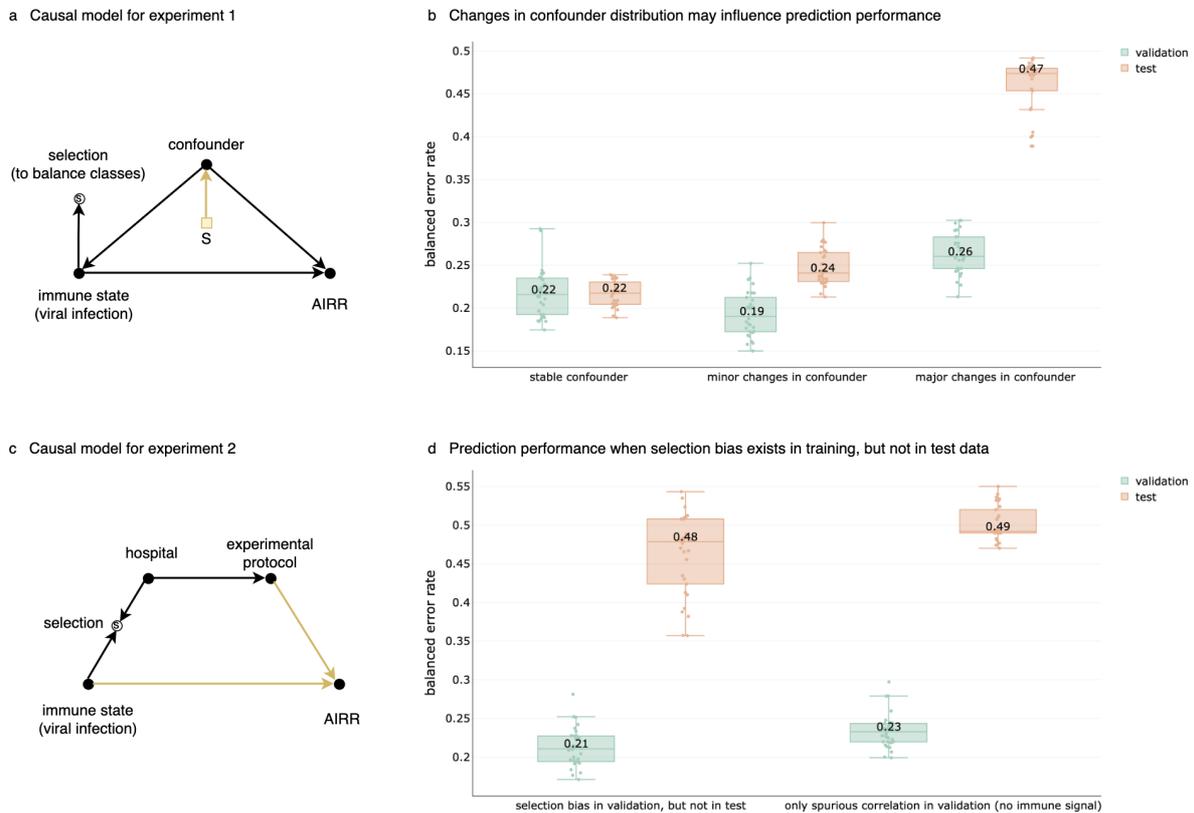

**Figure 4**: **Experiments showing immune state prediction performance under different causal models**. All results are shown for 500 AIRRs for training and validation and 500 for testing, with 500 TCRβs per repertoire, across 30 repetitions. Prediction performance was measured by balanced error rate (false positive / (true negative + false positive) + false negative / (true positive + false negative))/2), with median values shown in the plots. The immune signal present in diseased AIRRs consisted of 3-mers that were implanted approximately in the middle of the receptor sequence. For all scenarios (b, d), 3-mer frequencies and logistic regression were used, as previously described by Kanduri and colleagues[108]. In the subfigures (b, d), the median values are shown. **a**. Causal model with immune state, AIRR, and a confounder for Experiment 1. The yellow arrow denotes that the confounder distribution changes between the training (validation) and test population. The immune state is balanced in both populations. **b**. The balanced error rate when the training (validation) and test set originate from the same distribution (confounder distribution is stable), when the distribution changes slightly ($P(confounder)_{validation}=0.6$, $P(confounder)_{test}=0.8$), and when the confounder distribution substantially changes ($P(confounder)_{validation}=0.8$, $P(confounder)_{test}=0.01$). **c**. The causal model for Experiment 2 where the experimental protocol causes batch effects through the presence of selection bias. The edges in yellow from experimental protocol to AIRR and from immune state to AIRR denote the relations that are modified in the experiment. **d**. The balanced error rate when the selection bias is present in the training, but not in the test population, when there is an immune signal, and when no immune signal is present.

We also show that in the absence of any relation between the immune state and the AIRR, it is possible for an ML model to learn a spurious correlation because of the selection bias and achieve deceivingly good performance (depending on how strong the correlation between the immune state and hospital and protocol is). When a model trained on such data is applied to new data where the selection bias is not present, its performance is random (Figure 4d).

For both experiments, we simulated 500 AIRRs for training and 500 for testing using OLGA[109] with 500 TCRβ sequences per AIRR from the default human model. To simulate the immune state and the confounder from the causal graph, we used DagSim[110]. As the next step, we implanted 3-mers into some of the AIRs of individual repertoires with the appropriate immune state and confounder values. To assess ML performance, we encoded the AIRRs via 3-mer frequencies and fitted logistic regression with the L1 penalty to perform the prediction, as previously described by Kanduri and colleagues[108]. The model performance was measured via balanced error rate, defined as (false positive / (true negative + false positive) + false negative / (true positive + false negative))/2. For each setting, the analysis was repeated 30 times. For introducing signals into naive OLGA repertoires and for AIRR-ML analyses, we used immuneML[111].

# The complexities of batch effects in the AIRR setting as compared to classic molecular biomarkers

For classic molecular biomarkers, like gene expression biomarkers of cancer subtypes[112], it is well established that batch effects may influence analyses and should be accounted for[87]. The common assumption is that a particular batch will introduce a systematic shift of gene expression values. Since gene expression data are represented naturally as a numeric matrix, batch effects can be corrected for by regressing out the influence of batch by learning a linear regression from a batch index covariate to the log expression values across patients for each gene at a time[113].

For AIRR data, it is not equally obvious or established how datasets are influenced by batch effects. For example, batch effects are proposed to influence which receptors (based on their V/J genes) are preferentially captured by DNA sequencing[71]. Furthermore, AIRR data intrinsically come in the form of a set of sequences, and it is not even clear how the data should be represented when investigating batch effects. An additional challenge here is to select the representation to reflect both the biological signal and to allow for accounting of batch effects, where biological signal and batch effect might manifest differently.

To compare the feasibility of handling batch effects in AIRR data versus other high-throughput molecular datasets, we performed a simulation experiment on handling batch effects in AIRR versus transcriptomic data.

For the transcriptomic setting, we simulated RNA-seq count datasets with realistic batch and biological effects, where the biological condition correlated highly with batches. We assessed the performance of a predictive model in predicting the true biological condition of unseen test observations based on gene expression data in three different scenarios: (i) where no batch effects exist in the training data, (ii) where batch effects exist in the training data and are not accounted for, and (iii) where batch effects exist in the training data and handled using a method that regresses out the batch effects[113]. We repeated the experiments using ten independent datasets in each scenario to provide more robust estimates of the performance metrics. We observed that the performance of the predictive model on test data was close to perfect when no batch effects exist (control setting). In the case where the training data contained batch effects, the performance improved after regressing out the batch effects (Figure 5a). We noticed that in all three scenarios, all non-zero coefficients of the L1-regularized model reflected the true (introduced) signal (Figure 5b).

For illustrative purposes and clearer discussion, we did for the AIR setting focus on the prediction of individual receptor specificity instead of repertoire immune state. We simulated a dataset of 5000 immune receptor sequences that included the biological signal (k-mers implanted in the middle of the sequence to reflect epitope specificity) and batch effects (k-mers implanted at the beginning of the sequences to reflect gene biases). We compared the same three scenarios as in the transcriptomic setting: when there is no batch effect (control setting), when there is an uncorrected batch effect, and corrected batch effect. In cases where the batch effect was present, it was highly correlated with the receptor specificity in training and uncorrelated with receptor specificity in the test data. The simulation and analyses were repeated 5 times to establish performance bounds. We used a k-mer frequency encoding (k=3) as a representation for the receptors and chose to perform batch correction with linear regression and immune receptor specificity classification with logistic regression on this k-mer encoding of the data. In our simulations, while true immune-related signals were found also in scenarios with batch effects, these true signals were mixed alongside spurious signals when the biological signals were sufficiently subtle (Figure 5d). When trying to correct for batch effects by directly regressing out batch status on the k-mer representation, the error rate was reduced but was still higher compared to when batch effects were absent (Figure 5c).

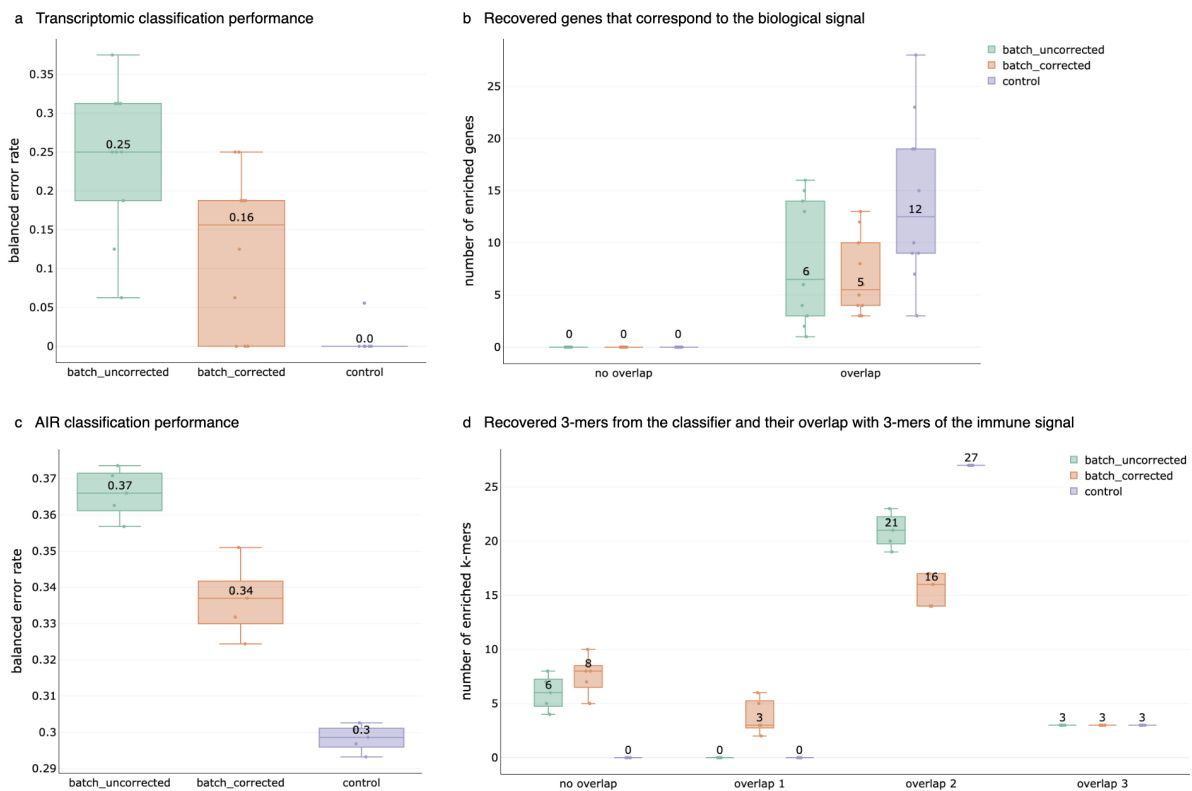

**Figure 5**: **Batch effects, both uncorrected and corrected through linear regression for each feature separately, may lead to higher error rates in transcriptomic and AIR settings, and might result in classifiers learning spurious signals, especially in the AIR setting**. The results are shown for three scenarios: batch effects present but not corrected, corrected batch effects using linear regression on k-mer frequencies, and no batch effects in the data (control scenario). The median values are shown in all the subfigures. **a.** Balanced error rate of predicting the disease state from gene expression across the three scenarios, where having uncorrected batch effects leads to the highest error rate. **b.** The number of enriched genes detected by the *L1*-regularized logistic regression model that overlap and do not overlap with the simulated biological signal. **c.** The balanced error rate of predicting receptor specificity across the three cases using k-mer frequencies for data representation and logistic regression, where having uncorrected batch effects lead to the highest error rate, although the difference is small. **d.** The recovered 3-mers from the logistic regression model are grouped by how much they overlap with the 3-mers of the immune signal, across the three scenarios. The recovered 3-mers from the model are obtained as the features corresponding to the 30 largest coefficients in the L1-regularized logistic regression model in absolute value.

While the findings in both AIRR and transcriptomic settings demonstrate that regressing out batch effects might improve the desired objective of high predictive performance (especially in the transcriptomic setting), it is at the same time worth noting that the degree of such improvement depends on the study design settings such as the degree of correlation between batches and biological condition, as well as the magnitude and type of influence of batch and biological effects.

# Conclusion

## Advice to AIRR researchers on study design and computational processing

To learn AIRR-based biomarkers that generalize well to clinical settings, we propose the following guidelines: (i) Ensure that batch effects, although nearly always present, only influence the observed AIRR and are not correlated with the immune state. To this end, use the same experimental protocols to sequence individuals with different immune states. If study participants are recruited or samples are processed in groups over time, ensure that each group includes both diseased and healthy individuals to minimize the impact of batch effects (Figure 4d). (ii) Internal validity, occurring when the targeted probability distribution is learned instead of noise (Focus Box 3), has to be achieved through appropriate assessment strategies and sufficiently large study cohorts[42]. Additionally, it is necessary to also recruit a sufficient number of participants for each of the confounder value groups. While this number may depend on the variability of the study cohort and the ML approach, when replicating the analysis by Emerson and colleagues[40], for example, we have previously found that even with a cohort size of 200 the classifier had close to random performance, while the model fitted on 400 individuals behaved similarly to the model estimated on the full dataset[111]. (iii) When recruiting study participants, avoid selection biases that may introduce spurious associations, as shown in Experiment 2 where the ML method learns the protocol instead of the immune state due to their correlation (Figure 4d). One exception to avoiding selection biases is when they are deliberately introduced (and compensated for) to enrich signals for machine learning, for example by balancing the classes when training a prediction model. Furthermore, in case the target population is known to differ from the source (training) population in a variable that has a major influence on AIRRs or immune state, as exemplified in Experiment 1 with confounders (Figure 4a, b), we advise using techniques such as inverse probability weighting[17], importance weighting[12], or importance sampling[114], to better reflect the expected distribution in the target environment (either at the study recruitment or analytical phase). We illustrate how failing to follow these recommendations might influence the prediction task in worked examples (Figure 4).

## Proposed reporting standards for AIRR diagnostic study design and computational analysis

To increase the trustworthiness of AIRR diagnostic studies and ensure their applicability in future use cases such as meta-analyses where multiple studies are examined together to provide a better answer to the research question, we propose the following reporting standards. (i) Report the sets of AIRR samples that have been processed together in batches. (ii) For each AIRR, provide information on recruitment source, experimental protocol, and institution. (iii) If external validity is anticipated, define the target (deployment) setting where the diagnostic could be applied. (iv) Report metadata, including information on sex, age, HLA, and similar properties as outlined by the MiAIRR standard[115,116]. Provide results per strata for any variable considered to have a major impact on AIRR and immune state (consult the state-of-the-art in the AIRR field and disease field at the time of publication). Include information on genetic ancestry and aim to cover diverse patient cohorts[57,117–119]. Additionally, reviewing study protocols in advance, e.g., through Registered Reports[120–122], may alleviate some of the concerns described in previous sections, especially related to the researcher's degree of freedom in assessing internal validity[123].

## Suggested research directions for the AIRR field

The many uncertainties discussed in this paper regarding how to define the relevant causal graph for AIRR-based diagnostic settings point to several open research questions for the AIRR field.

A major open question is the degree to which the HLA allotypes influence AIRRs[5,6,40,70,124]. Strong correlations between HLA and CDR3 regions of TCRs have recently been observed indicating that HLA risk allotypes might increase the frequency of autoreactive TCRs already during T cell development[70]. From a diagnostic perspective, the HLA influence can be seen as two sub-questions: (i) the degree to which HLA leaves a detectable mark in the overall AIRR that can be leveraged to capture the disease-predictive information of HLA by AIRR sequencing alone (leverage the indirect path AIRR ← HLA → disease), and ii) the degree to which HLA moderates the direct AIRR–disease relation so that machine learning models need to rely on (learn) distinct predictive patterns for individuals with different HLAs.

Similarly, the degree to which genetics leaves detectable marks on AIRR repertoires would determine whether AIRR-based diagnostics could exploit disease-predictive information inherent in the reflected genetic variables. A related question is the degree to which the information captured by AIRR-based diagnostics is overlapping with or is complementary to the predictive information provided by explicit full-genome genotyping or HLA haplotyping.

Another challenge comes from AIRRs being complex and high-dimensional. It is currently unclear to what extent different biases discussed in previous sections are detectable in the high-dimensional data and how they influence the classification task. Additionally, the many mechanistic ways in which disease states and AIRRs influence each other across time points should be examined in greater detail to see the impact on the diagnostic models.

To illustrate the current AIRR diagnostics challenges in light of causal modeling, we relied on simulated data, since the ground truth in the form of rules guiding the (AIRR) data generation is generally unknown. In these cases, simulation tools such as OLGA[109], immuneSIM[125], or Absolut![126] for simulating AIRRs, and DagSim[127] for simulating data from causal graphs are extremely useful for benchmarking different ML approaches because they provide fully annotated data. Equipped with the understanding from simulations, we might be able to better understand and analyze experimental data, build generalizable diagnostics, and at least partially learn the true causal model of biological processes.

While we argue causality is important for ML robustness and diagnostic development study design, causality is also an aim in itself in terms of describing how the adaptive immune system works, as also pointed out by Bizzarri and colleagues for the general case of cell biology[128]. Establishing a causal AIRR model would enable not only the improvement of AIRR-based diagnostics but would also provide causal interpretations and estimate the effects of interventions. For example, a sufficiently detailed AIRR model, along with the methodology to successfully handle high-dimensional data, may allow computational screening of new candidate therapies[129] and vaccination procedures[130].

**Data availability**

All data and results of the analyses presented in the manuscript are openly available on Zenodo:

- experiment 1 (https://zenodo.org/record/7756163),
- experiment 2 (https://zenodo.org/record/7752837),
- experiment 3 (https://zenodo.org/record/7752115, https://zenodo.org/record/7727894).

**Code availability**

All code used for the analyses in the manuscript is publicly available on GitHub: https://github.com/uio-bmi/CausalAIRR.

**Acknowledgments**

We acknowledge generous support by The Leona M. and Harry B. Helmsley Charitable Trust (grant number 2019PG-T1D011, to V.G.), the UiO World-Leading Research Community (to V.G. and L.M.S.), the UiO:LifeScience Convergence Environment Immunolingo (to V.G. and G.K.S.), the UiO:LifeScience Convergence Environment RealArt (to G.K.S. and C.K.), EU Horizon 2020 iReceptorplus (grant number 825821, to V.G. and L.M.S.), a Research Council of Norway FRIPRO project (grant number 300740, to V.G.),  a Norwegian Cancer Society Grant (#215817, to VG), a Research Council of Norway IKTPLUSS project (grant number 311341, to V.G. and G.K.S.), and Stiftelsen Kristian Gerhard Jebsen (K.G. Jebsen Coeliac Disease Research Centre, to L.M.S. and G.K.S.).

**Contributions**

M.P., V.G., and G.K.S. conceived the study. M.P., G.S.A.H., and C.K. performed the experiments. J.P., M.W., and L.M.S., C.K. provided critical feedback. M.P., V.G., and G.K.S. drafted the manuscript. G.K.S. supervised the project. All authors read and approved the final manuscript and are personally accountable for its content.

**Competing interests**

V.G. declares advisory board positions in aiNET GmbH, Enpicom B.V,  Absci, Omniscope, and Diagonal Therapeutics. VG is a consultant for Adaptyv Biosystems, Specifica Inc, Roche/Genentech, immunai, and LabGenius.